\documentclass[11pt]{article}
\usepackage{amsmath,amssymb,color,graphics,epsfig,cite}

\usepackage{amsfonts}

\newcommand{\hoch}[1]{$\, ^{#1}$}

\newcommand{\be}{\begin{equation}}
\newcommand{\ee}{\end{equation}}
\newcommand{\bea}{\setlength\arraycolsep{2pt} \begin{eqnarray}}
\newcommand{\eea}{\end{eqnarray}}
\newcommand{\nn}{\nonumber}

\def\ft#1#2{{\textstyle{\frac{\scriptstyle #1}{\scriptstyle #2} } }}
\def\fft#1#2{{\frac{#1}{#2}}}

\def\0{{\sst{(0)}}}
\def\1{{\sst{(1)}}}
\def\2{{\sst{(2)}}}
\def\3{{\sst{(3)}}}
\def\4{{\sst{(4)}}}
\def\5{{\sst{(5)}}}
\def\6{{\sst{(6)}}}
\def\7{{\sst{(7)}}}
\def\8{{\sst{(8)}}}
\def\sst#1{{\scriptscriptstyle #1}}

\usepackage{xcolor}

\def\si{{\psi}}

\begin{document}

\begin{center}
{\Large {\bf On the Lagrangian Holographic Relation \\
at $D\rightarrow2$ and $4$ Limits of Gravity}}

\vspace{20pt}
{\large H. Khodabakhshi\hoch{1}, H. L\"u\hoch{1} and R.~B.~Mann\hoch{2,3}}

\vspace{10pt}

{\it \hoch{1}Center for Joint Quantum Studies and Department of Physics,\\ School of Science, Tianjin University,\\ Yaguan Road 135, Jinnan District, Tianjin 300350, China}

{\it \hoch{2}Department of Physics and Astronomy, University of Waterloo, \\Waterloo, Ontario, Canada, N2L 3G1}

{\it \hoch{3}Perimeter Institute, 31 Caroline St. N., Waterloo, Ontario, N2L 2Y5, Canada
}
\vspace{40pt}

\underline{ABSTRACT}
\end{center}

The gravitational Lagrangian can be written as a summation of a bulk and a total derivative term. For some theories of gravity such as Einstein gravity, or more general Lovelock gravities, there are Lagrangian holographic relations between the bulk and the total derivative term such that the latter is fully determined by the former.  However at the $D\rightarrow 2\&4$ limit, the bulks of Einstein or Gauss-Bonnet theories become themselves total derivatives. Performing the Kaluza-Klein reduction on Einstein and Gauss-Bonnet gravities gives rise to some two-dimensional or four-dimensional scalar-tensor theories respectively. We obtain the holographic relations for the $D = 2$ and $D = 4$ cases, which have the same form as the holographic relations in pure gravity in the foliation independent formalism.

\vfill{\footnotesize  h\_khodabakhshi@tju.edu.cn\ \ \ mrhonglu@gmail.com  \ \ \ rbmann@uwaterloo.ca}



\thispagestyle{empty}
\pagebreak



\section{Introduction}

The gravitational Lagrangian must be a scalar function of the metric and   at least its first and  second derivatives. Any covariant Lagrangian that  contains the metric and its first derivatives only will be a trivial constant \cite{Pad1}. We also know that for some Lagrangians constructed from the specific invariant polynomial combinations of the Riemann tensor, even at the higher orders, the field equations remain second order in the derivatives of the metric. These include the Einstein, Lovelock \cite{Love} and Horndeski \cite{Hor} gravities for general metrics, and even some quasitopological gravity theories \cite{GQT1, GQT2} in some special classes of the metric ansatz. In fact for these types of Lagrangians, the field equations are effectively derived from a first-order sub-Lagrangian that is hidden within the whole Lagrangian involving total derivatives. Hence one could write these types of Lagrangians as a summation of a bulk and a total derivative term as
\be
 \sqrt{-g} L_{G}( g,\partial g, \partial^2 g)=\sqrt{-g} L_{\rm bulk}+\partial_\mu (\sqrt{-g}J^\mu)\,,
\ee
to obtain the hidden first-order Lagrangian \cite{TB, FOL, Pad2}. In this parsing of the Lagrangian, the first term, which is not covariant, determines completely the covariant equations of motion. Thus, the total derivative term is such that   covariance can be established at the level of Largangian. It turns out that total derivative term is completely determined by the bulk term, giving rise to the Lagrangian holographic relation \cite{TB, FOL, Pad2}. This phenomenon has recently been shown to be true for   Gauss-Bonnet gravity, or general Lovelock gravities of any order $k$, but with some additional subtleties \cite{HL1}. In the AdM formalism, where one coordinate, say $r$, is treated as special, the bulk Lagrangian does not involve derivatives of $r$ more than the first order. In the foliation independent approach, on the other hand, the bulk Lagrangian can still have second derivatives of the metric.  Nevertheless in both approaches, there exists a Lagrangian holographic relation where the total derivative terms are not arbitrary, but determined completely by the bulk \cite{HL1}. These holographic relations may provide a deeper insight into several aspects of classical and semiclassical gravity that have no explanation in the conventional approach \cite{Pad1,Pad2,HL1}.

In the formalism given in \cite{HL1}, the Lagrangian holographic relation for Lovelock gravity of any order $k$ is formally valid in all dimensions. This leads to trivial cases that turn out to be interesting. In $D=2k$ dimensions, the total derivative in   Lovelock gravity is determined by the bulk term that is itself a total derivative. However for $k=2$ Glavan and Lin \cite{Lin} recently proposed that there might exist a $D\rightarrow 4$ limit of Gauss-Bonnet gravity by rescaling the coupling constant of the theory according to
\begin{equation}
\alpha \rightarrow \fft{\alpha}{D-4}\,,\label{limit}
\end{equation}
and taking the $D\rightarrow4$ limit. One concrete realization of this proposal is to consider the Kaluza-Klein reduction of Gauss-Bonnet gravity in general dimensions to four-dimensions to obtain the theory with an additional scalar field. Then there exists a consistent and smooth limit of the four-dimensional theory under \eqref{limit} \cite{LP,TK,Hennigar:2020lsl,RR}. This generalizes a method employed by Mann and Ross to obtain a limit of the Einstein gravity in $D = 2$ dimensions \cite{RB1,Hennigar:2020lsl,Grumiller:2007wb}.

In this paper we consider the holographic Lagrangian relations in the Einstein and GB gravities at their respective $D\rightarrow 2$ and $D\to 4$ limits. In section 2, we briefly review the Lagrangian holographic relation in Einstein gravity. We then consider the $D\rightarrow 2$ limit that leads to the 2-dimensional scalar-tensor theory. Then using the obtained holographic relation for Einstein gravity in $D$ dimensions, we will   write the holographic relation for $D=2$   by fixing the arbitrary coefficients of the total derivative terms.  Since any two-dimensional manifold is conformally flat,  we can construct the holographic relation in this special case. In section 3, we reconsider the analogous problem for the Gauss-Bonnet Lagrangian.  We summarize our work in a concluding section.

\section{The $D\rightarrow 2$ limit of gravity}

In this section, we consider Einstein gravity in  the $D\rightarrow 2$ limit \cite{RB1,Grumiller:2007wb,Hennigar:2020lsl} and obtain the 2-dimensional scalar-tensor theory. From this we shall obtain the Lagrangian holographic relation.

\subsection{Holographic relation in general $D$ dimensions}

We begin with a brief review of Einstein gravity in general $D$ dimensions \cite{Pad1, Pad2, HL1}. The Lagrangian for the Einstein-Hilbert can be split into a bulk term, which is a quadratic polynomial function of the affine connection, and a total derivative term, namely
\be\label{HEH}
{\cal L}_{\rm EH}=\sqrt{-g} R = \sqrt{-g} L^{\rm E}_{\rm bulk}(\Gamma) + \partial_\mu (\sqrt{-g} J_{\rm E}^{\mu}),
\ee
where
\bea
L^{\rm E}_{\rm bulk}(\Gamma)&=& -2 \delta^{\mu_1\mu_2}_{\nu_1\nu_2} (\Gamma^2)^{\nu_1\nu_2}_{\mu_1\mu_2}\equiv -2 \delta^{\mu_1\mu_2}_{\nu_1\nu_2} \Gamma^{[\nu_1}{}_{\alpha [\mu_1} \Gamma^{|\alpha|}{}_{\mu_2]\beta} g^{\nu_2]\beta}\,,\nn\\
J_{\rm E}^{\mu_1} &=& 2 \delta^{\mu_1\mu_2}_{\nu_1\nu_2} g^{\rho \nu_2} \Gamma^{\nu_1}{}_{\rho\mu_2}=g^{\rho\sigma} \Gamma^{\mu_1}{}_{\rho\sigma} - g^{\rho\mu_1}\Gamma^\sigma{}_{\rho\sigma} \,.\label{einsteinbulksurf}
\eea
The theory satisfies a holographic relation insofar as   the surface term is determined by the bulk Lagrangian via
\be
J^{\mu}_{\rm E} = -\delta^\sigma_\rho \fft{\partial L^{\rm E}_{\rm bulk}}{\partial \Gamma^\sigma{}
	_{\mu\rho}}\,.\label{holorelation1}
\ee
where $\Gamma^\sigma{}_{\mu\rho}$ and $g_{\mu\nu}$ are not independent quantities. Indeed it can be established that dynamics of the Einstein-Hilbert action is determined by the bulk Lagrangian fully, via the variational principle, namely
\be
\delta (\sqrt{-g} L^{\rm E}_{\rm bulk}(\Gamma)) =
G_{\mu\nu} \delta g^{\mu\nu} + \partial_{\mu_1} \Big(\sqrt{-g} \big(g_{\mu\nu} J_{\rm E}^{\mu_1} - 2 \delta^{\mu_1\mu_2}_{\nu_1\nu} \Gamma^{\nu_1}{}_{\mu\mu_2}\big) \delta g^{\mu\nu}\Big)\,,
\ee
where $G_{\mu\nu}=R_{\mu\nu} - \fft12 g_{\mu\nu} R$ is the Einstein tensor  and the variation is with respect to the metric.

It is important to note that $J^\mu$ here is not a tensor, since it involves the undifferentiated affine connection $\Gamma$. Consequently neither the bulk nor the surface terms are separately invariant under a general coordinate transformation. This implies that for any special class of metric ansatz for a given coordinate choice, the total derivative term cannot be simply discarded from the Lagrangian to study the classical dynamics of the system. As a concrete example, we consider the most general static and spherically symmetric ansatz in four dimensions
\be
ds^2 = - h(r) dt^2 + \fft{dr^2}{f(r)} + r^2 \Big(\fft{d\chi^2}{1-\chi^2} + (1-\chi^2) d\phi^2\Big)\,.\label{metricansatz}
\ee
It is easy to verify that
\bea
\sqrt{-g} L_{\rm bulk}&=&2\sqrt{\fft{f}{h}} (rh)'\,,\qquad J^t=0=J^\phi\,,\nn\\
\sqrt{-g} J^r &=& - \sqrt{\fft{f}{h}} r (4h + r h')\,,\qquad
\sqrt{-g} J^\chi = 2\chi \sqrt{\fft{h}{f}}\,.
\eea
Thus we see that while the term $\partial_r (\sqrt{-g} J^r)$ does not contribute to the equations of motion, the term $\partial_\chi (\sqrt{-g} J^\chi)$, which depends on $r$ only, does. This is because, as we mentioned earlier, $J^\mu$ is not gauge covariant. In particular, the $J^\chi$ component is not well defined in the spherically-symmetric ansatz \eqref{metricansatz}, since $J^\chi J_\chi$ diverges at both north and south poles ($\chi=\pm1$).

The holographic relation (\ref{holorelation1}) works in all dimensions, but it becomes trivial in $D\le 2$.  In particular, $\sqrt{-g}L^{\rm E}_{\rm bulk} (\Gamma)$ is itself a total derivative in $D=2$ such that $G_{\mu\nu}$ vanishes identically. However, when the Newton's constant vanishes in $D=2$, there exists a nontrivial limit such that the theory reduces to a scalar tensor theory.  We shall discuss this next.

\subsection{Scalar tensor theory as $D\rightarrow 2$ limit of gravity}

We begin with Einstein gravity in general $D=n+2$ dimensions, reduced on some $n$-dimensional Einstein space, keeping only the scalar breathing mode $\phi$. The reduction ansatz is \cite{Grumiller:2007wb}
\be
d\hat s_{n+2}^2 = ds_2^2 + e^{2\phi} d\Sigma_{n,\lambda}^2\,,
\ee
where internal metric $d\Sigma_{n,\lambda}^2$ is Einstein with cosmological constant curvature $\lambda$:
\be
\tilde R_{\mu \nu}=(n-1) \lambda \tilde g_{\mu \nu}\,.
\ee
Thus we have
\be
\fft{1}{16\pi \hat G}  \int d^{n+2}x\sqrt{-\hat g}\hat R =
\fft{1}{16\pi G} \int d^2x \sqrt{-g} e^{n\phi} \Big(R + n(n-1)\lambda e^{-2\phi} - 2n\Box\phi - n(n+1)(\nabla\phi)^2\Big)\,,
\ee
where $G=\fft{\hat G \sqrt{\lambda^n}}{ \omega_{n}}$, and $\omega_n$ is the internal space volume of unit cosmological constant . Integrating by part gives the two-dimensional Lagrangian
\be
{\cal L}_2^{(n)}=\sqrt{-g} \bigg( e^{n\phi} \Big(R+n(n-1)\lambda e^{-2\phi} + n(n-1) (\nabla\phi)^2\Big) -2n  \nabla_\mu (e^{n \phi} \nabla^\mu \phi)\bigg) .
\ee
Now if we simply take $n=0$, we have
\be
{\cal L}_2^{(0)}=\sqrt{-g} R\,,
\ee
which is a total derivative.  However, if the Newton's constant also vanishes in this limit, e.g.~$\hat G\rightarrow \hat G n\rightarrow 0$, we can obtain a nontrivial limit, namely
\be
\lim_{n\rightarrow 0} \fft{1}{n}\Big({\cal L}_2^{(n)}-{\cal L}_2^{(0)}\Big)=
\sqrt{-g} \Big(\phi R -\lambda e^{-2\phi} - (\nabla\phi)^2- 2 \nabla_\mu (\nabla^\mu \phi)\Big).
\ee
We therefore obtain a nontrivial scalar-tensor theory as the $D\rightarrow 2$ limit of Einstein gravity, with the Lagrangian
\be \label{EH2}
{\cal L}_2 = \sqrt{-g} \Big(\phi R -\lambda e^{-2\phi} - (\nabla\phi)^2 +a \nabla_\mu (\nabla^\mu \phi)\Big),
\ee
where $a=-2$. For $\lambda=0$ the $R=T$ theory is recovered \cite{RB1}.
Note that performing the $n\rightarrow 0$ limit after the reduction yields a surface term $\sqrt{-g} \Box \phi$ that is typically discarded. This term is very different from our earlier surface term in Einstein gravity since $\nabla^\mu \phi$ is a proper tensor. This allows us to modify the constant coefficient $a$ at will without sacrificing the covariance or altering the classical dynamics even for a specialized but consistent ansatz.

\subsection{Holographic relations}

\subsubsection{Conformal characterization}
\label{subsection boost solution a}

We now study the holographic relation associated with the two-dimensional scalar-tensor theory \eqref{EH2}. It is well-known that every two-dimensional manifold $(\mathcal{M}, g)$ is conformally flat. We therefore consider a generic two-dimensional metric in the conformally flat frame
\be
ds_2^2=\Omega (x,y) ^2(-dx^2+dy^2),
\ee
where $\Omega (x,y)=e^{\si (x,y)} > 0$ is the conformal factor and it is a smooth function. The
Ricci scalar curvature $R$ can be expressed as a function of $\Omega$ and its partial derivatives by
\be
R= \frac{2 (\nabla \Omega)^2-2 \Omega \nabla_\mu (\nabla^\mu \Omega)}{\Omega^2}=-2\nabla_\mu (\nabla^\mu \si).
\ee
Substituting the above equation into Eq.~(\ref{EH2})  and integrating by parts  yields
\be\label{X}
{\cal L}_2 = \sqrt{-g} \Big(2 \nabla^\mu \si \nabla_\mu \phi -\lambda e^{-2\phi} - (\nabla\phi)^2 + \nabla_\mu \big(-2 \phi \nabla^\mu \si + a \nabla^\mu \phi\big) \Big).
\ee
Choosing $a=0$, we find the following holographic relation
\be \label{Holo}
{\cal L}_2 = \sqrt{-g}L_{\rm bulk} - \partial_\mu \Big(\phi \frac{\partial (\sqrt{-g} L_{\rm bulk})}{\partial (\nabla_\mu \si)} +\phi \frac{\partial (\sqrt{-g} L_{\rm bulk})}{\partial (\nabla_\mu \phi)} \Big),
\ee
where the bulk term is
\be \label{Holob}
L_{\rm bulk} = 2 \nabla^\mu \si \nabla_\mu \phi -\lambda e^{-2\phi} - (\nabla\phi)^2.
\ee

\subsubsection{Foliation independent relation}\label{subsection boost solution b}

Substituting the Ricci scalar from
 (\ref{HEH}) into  (\ref{EH2}) and integrating by parts gives
\be \label{EH2H}
{\cal L}_2 = \sqrt{-g} \Big(\phi L_{\rm bulk}^{\rm E}-J^{\mu}_{\rm E} \nabla_\mu \phi -\lambda e^{-2\phi} - (\nabla\phi)^2 +\nabla_\mu \big(\phi J^{\mu}_{\rm E} +a\nabla^\mu \phi\big) \Big).
\ee
We find that for $a = -\ft12$, the theory satisfies the following holographic relation
\be\label{d2holorelation}
{\cal L}_2 = \sqrt{-g} L_{\rm bulk} - \partial_\mu \Big(\delta_\beta ^\alpha \frac{\partial(\sqrt{-g} L_{\rm bulk})}{\partial \Gamma^\alpha {}_{\mu \beta}}\Big),
\ee
with
\be \label{bhol}
L_{\rm bulk}=\phi L_{\rm bulk}^{\rm E}-J_{\rm E}^{\mu} \nabla_\mu \phi -\lambda e^{-2\phi} - (\nabla\phi)^2\,,
\ee
where $L_{\rm bulk}^{\rm E}$ and $J_{\rm E}^\mu$ are given in \eqref{einsteinbulksurf}. In the above, we made use of the identity
\be
\delta_\alpha^\beta \frac{\partial J^{\nu}_{\rm E} }{\partial \Gamma^\alpha {}_{\mu \beta}}=-\ft12(D-1) g^{\mu\nu}
=-\ft12  g^{\mu\nu}
\,.\label{identity1}
\ee
It is intriguing to observe that although the nontrivial theory in $D=2$ is the scalar-tensor theory with a scalar field $\phi$, the holographic relation \eqref{d2holorelation} takes the same form as the one of the pure and trivial $D=2$ gravity, as given by \eqref{holorelation1}.

\section{$D\rightarrow 4$ limit of GB gravity}

A similar approach can be applied to the $D\rightarrow 4$ limit of GB gravity. The relevant Lagrangian is given by
\be
{\cal L}_D = \alpha \sqrt{-g} {\mathcal G}\,,\qquad \mathcal{G}=R_{\mu \nu \rho \sigma} R^{\mu \nu \rho \sigma}-4R_{\mu \nu} R^{\mu \nu}+R^2\,.
\ee
For fixed coupling constant $\alpha$, the theory is trivial at $D=4$ since the Lagrangian is a total derivative. However, nontrivial limits of solutions to the equations of motion can be obtained when the coupling constant becomes divergent, with $\alpha\rightarrow \alpha/(D-4)$ \cite{Lin}. One way to take this limit is to treat it as a special class of scalar-tensor Horndeski gravity \cite{Hor}, obtained from the Kaluza-Klein reduction, taking the dimension of internal space to be zero \cite{LP, TK, Hennigar:2020lsl,RR}. In this approach, one begins with the Kaluza-Klein ansatz
\be
d\hat{s}^2_{n+4}=ds_4^2+e^{2\phi} d\Sigma_{n,\lambda}^2\,,
\ee
where $d\Sigma_{n}^2$ to be the line element of $n$-dimensional maximally-symmetric space with constant $\lambda$, and $\phi$ is the breathing mode depending only on the external $4$ dimensional coordinates. The Gauss-Bonnet combination becomes
\begin{align}
\hat{\mathcal{G}}=&\mathcal{G}+4 n (n+1) G^{\mu \nu} \nabla_{\mu} \phi \nabla_{\nu} \phi+4n (n-1) (n+1) \Box \phi (\nabla \phi)^2+n(n-1) (-2+n(n-1)) (\nabla \phi)^4\nonumber\\
&
+2 n(n-1) e^{-2\phi} \lambda R+n (n-1) (n-2) (n-3) e^{-4\phi} \lambda^2-2 n(n-1) (n-3) e^{-2\phi} \lambda (\nabla \phi)^2
\nonumber\\
& + \nabla_\mu \bigg(8n G^{\mu \nu} \nabla_\nu \phi -4n (n-1) \nabla^{\mu} \phi (\nabla \phi)^2 +4 n(n-1) \big(\Box \phi \nabla^{\mu} \phi - (\nabla^\mu \nabla^\nu \phi) (\nabla_\nu \phi) \big)
\nonumber\\
&
-n(n-1) (n-3) e^{-2\phi} \lambda \nabla^{\mu} \phi\bigg) ,
\end{align}
where $G^{\mu \nu}$ is the Einstein tensor. It can be easily seen that when $n=0$, the combination $(\hat{\mathcal{G}}-\mathcal{G})/n$ has a nontrivial limit. This leads to a four-dimensional   Horndeski-type of scalar-tensor theory \cite{LP}
\begin{align} \label{GB4surface}
{\cal L}_{4}= &\sqrt{-g} \Big\{\phi \mathcal{G}+4 G^{\mu \nu} \nabla_{\mu} \phi \nabla_{\nu}\phi - 4 \Box \phi (\nabla \phi)^2+2 (\nabla \phi)^4-2\lambda R e^{-2\phi}-12 \lambda (\nabla \phi)^2 e^{-2\phi} -6 \lambda^2 e^{-4\phi}
\nonumber\\
&+\nabla_{\mu}\bigg( a_1 G^{\mu \nu} \nabla_{\nu} \phi +a_2 (\nabla \phi)^2 \nabla^{\mu} \phi +a_3 \Box \phi \nabla^{\mu}\phi +a_4(\nabla^\mu \nabla^\nu \phi) (\nabla_\nu\phi)+a_5 e^{-2\phi} \lambda \nabla^{\mu} \phi  \bigg) \Big\}
\end{align}
where we have also included the total derivative terms, with $a_1=8$, $a_2=a_4=4=-a_3$ and $a_5=-3$.  As in the case of the scalar-tensor theory in $D=2$, the total derivative term in Eq.~(\ref{GB4surface}) is a proper tensor and hence the values of these coefficients will not affect the dynamics of the Lagrangian.

\subsection{Holographic relation}

The foliation independent holographic relation for GB or more general Lovelock gravities were obtained in \cite{HL1}. For the GB theory in general dimensions, one has
\bea\label{GBbulksurf}
\sqrt{-g} \mathcal{G} &=& \sqrt{-g} L^{\mathcal{G}}_{\rm bulk} +\partial_\mu (\sqrt{-g} J^{\mu}_{\mathcal{G}})\,,\nn\\
L^{\mathcal{G}}_{\rm bulk}(\Gamma,\partial\Gamma)&=&-4!\bigg(\, \delta^{\mu_1\mu_2\mu_3\mu_4}_{\nu_1\nu_2\nu_3\nu_4}\, (\Gamma^2)^{\nu_1\nu_2}_{\mu_1\mu_2}\, (\Gamma^2)^{\nu_3\nu_4}_{\mu_3\mu_4} +\, \delta^{\mu_1\mu_2\mu_3\mu_4}_{\nu_1\nu_2\nu_3\nu_4}\, (D\Gamma^2)^{\nu_1\nu_2}_{\mu_1\mu_2}\, (D\Gamma^2)^{\nu_3\nu_4}_{\mu_3\mu_4}\bigg)\,,\nonumber\\ J^{\mu}_{\mathcal{G}}&=&4!\,\delta^{\mu\mu_2\mu_3\mu_4}_{\nu_1\nu_2\nu_3\nu_4}\, \Gamma^{\nu_1}\!{}_{\mu_2}\!{}^{\nu_2} \,\Big((\Gamma^2)^{\nu_3\nu_4}_{\mu_3\mu_4} + (D\Gamma^2)^{\nu_3\nu_4}_{\mu_3\mu_4}\Big),
\eea
in which we have
\be
(D\Gamma^2)^{\nu_1\nu_2}{}_{\mu_1\mu_2} = \partial_{[\mu_1} \Gamma^{[\nu_1}{}_{\mu_2]\alpha} g^{\nu_2]\alpha}\,.\label{Gammasq}
\ee
The holographic relation is expressed as \cite{HL1}
\be
J^{\mu}_{\mathcal{G}} = - \delta^\nu_\rho \fft{\partial ( L^{\mathcal{G}}_{\rm bulk})}{\partial \Gamma^\nu{}_{\mu\rho}} - \Gamma^\nu{}_{\rho\sigma} \fft{\partial ( L^{\mathcal{G}}_{\rm bulk})}{\partial (\partial_\mu\Gamma^\nu{}_{\rho\sigma})}\,.\label{GBholorelation}
\ee
In other words, the surface term is completely specified by the bulk action.

Since GB gravity in $D=4$ is trivial, we now study the  holographic relation for the scalar tensor theory \eqref{GB4surface}. The identity
\be
\nabla_\nu \Box \phi -\nabla^\mu  \nabla_\nu \nabla_\mu \phi = -\nabla_\mu \phi R^\mu_\nu\,,
\ee
can be used to manipulate the $R^{\mu\nu}\nabla_\mu\nabla_\nu$ term in the Lagrangian \eqref{GB4surface}.
Substituting the Ricci scalar from~\eqref{HEH} and the Gauss-Bonnet term  $\mathcal{G}$ from~(\ref{GBbulksurf}) into (\ref{GB4surface}), and integrating by parts, we obtain
\begin{equation}\
\mathcal{L}_4= \sqrt{-g} L_{\rm  bulk} +\partial_\mu (\sqrt{-g} J^{\mu})\,,\label{GBbulksurf4}
\end{equation}
where
\bea
L_{\rm bulk}&=&\phi L_{\rm bulk}^{\mathcal{G}} - J^{\mu}_{\mathcal{G}} \nabla_{\mu} \phi -2\big((\nabla \phi)^2-\lambda e^{-2\phi} \big)L_{\rm bulk}^{\rm E} +2 J^{\mu}_{\rm E} \nabla_{\mu} \big((\nabla \phi)^2-\lambda e^{-2\phi} \big)
\nonumber\\
&&-4(\nabla \nabla \phi)^2+4 (\Box \phi)^2+2(\nabla \phi)^4-12 \lambda (\nabla \phi)^2 e^{-2\phi} -6 \lambda^2 e^{-4\phi}\,,\label{GBbulk} \\
J^{\mu}&=&\phi J^{\mu}_{\mathcal{G}}-2J^{\mu}_{\rm E}\big( (\nabla \phi)^2 -\lambda e^{-2\phi}\big) +(a_3-4) \nabla^{\mu}\phi \Box \phi +(4+a_4)  (\nabla^\mu \nabla^\nu \phi)(\nabla_{\nu}\phi)
 \nonumber\\
 &&+a_1 G^{\mu \nu} \nabla_{\nu} \phi +a_2 (\nabla \phi)^2 \nabla^{\mu} \phi+a_5 e^{-2\phi} \lambda \nabla^{\mu} \phi\,.\label{GBsurface}
\eea
We find that for the coefficients $a_1=4!$, $a_3=4$, $a_4=2$, $a_2=0$ and $a_5=6$ the total derivative term in Eq.~(\ref{GBbulksurf4}) can be written in terms of the bulk term as
\begin{align}
J^{\mu}=&- \delta^\nu_\rho \fft{\partial (L_{\rm bulk})}{\partial \Gamma^\nu{}_{\mu\rho}} - \Gamma^\nu{}_{\rho\sigma} \fft{\partial ( L_{\rm bulk})}{\partial (\partial_\mu\Gamma^\nu{}_{\rho\sigma})},
\end{align}
which is exactly the same form as \eqref{GBholorelation}. In the above derivation, we made use of the identity (\ref{identity1}), together with the following new identity
\begin{align}
	\delta^\nu_\rho \fft{\partial J^{\nu}_{\mathcal{G}}}{\partial \Gamma^\nu{}_{\mu\rho}} + \Gamma^\gamma{}_{\rho\sigma} \fft{\partial J^{\nu}_{\mathcal{G}}}{\partial (\partial_\mu\Gamma^\gamma{}_{\rho\sigma})}=-\frac{4!}{4} g^{\mu \gamma} g^{\nu_4 \nu'} \delta^{\sigma \nu \mu_3 \mu_4}_{\sigma \gamma \nu_3 \nu' }R^{\nu_3}{}_{\nu_4 \mu_3 \mu_4 }=4! (D-3) G^{\mu \nu}.
\end{align}

\section{Conclusion}

In this paper, we obtained the holographic Lagrangian relations in   Einstein gravity and in GB gravity in the $D\rightarrow 2$ and $D\to 4$ limits respectively. For pure gravity, Einstein, GB or more general Lovelock gravities all have a dimension independent holographic relation in the foliation independent formalism \cite{HL1}. However, in relevant dimensions, these theories become total derivatives and hence trivial. Using   Kaluza- Klein reduction, Einstein and GB theories can be written as scalar-tensor theories of the Horndeski type \cite{LP, TK,Hennigar:2020lsl,RR}. These theories involve total derivatives of tensor fields that are typically discarded since they do not affect either the covariance or the dynamics. We find that with suitable coefficients, we can establish the Lagrangian holographic relations for these scalar-tensor theories in the foliation independent formalism  \cite{HL1}. This fills the gap for gravitational theories that might be otherwise considered trivial.

The existence of some arbitrary total derivative terms in a scalar tensor theory that do not affect the dynamics nor the covariance shows that such a theory is not as tight as Einstein gravity. Although the dynamics of Einstein gravity is completely determined by the bulk Lagrangian, the additional covariance requirement makes the Einstein-Hilbert Lagrangian the unique theory. Such a requirement becomes less stringent  when matter or higher derivatives are involved. For example we can add $\sqrt{-g} \Box\phi$ to the Lagrangian with no effect when a scalar field is involved; or we can add $\sqrt{-g}\, \Box R$ to fourth-order derivative gravity without altering the dynamics or violating the covariance. The Lagrangian holographic relations thus provide a strong selection criterion.  This selection principle coincides with the covariance requirement in Einstein pure gravity, but its general physical interpretation requires further investigation.

\section*{Acknowledgement}

This work was supported in part by NSFC (National Natural Science Foundation of China) Grants No.~11875200 and No.~11935009 and by the Natural Sciences and Engineering Research Council of Canada.


\begin{thebibliography}{99}

\bibitem{Pad1}{T.~Padmanabhan, ``Gravitation: foundations and frontiers'' (Cambridge University Press, Cambridge, 2010).}

\bibitem{Love}{D.~Lovelock, ``The Einstein tensor and its generalizations,'' J. Math. Phys. 12 498 (1971) doi:10.1063/1.1665613.}

\bibitem{Hor}{G.W.~Horndeski, ``Second-order scalar-tensor field equations in a four-dimensional space,'' Int. J. Theor. Phys. 10 363 (1974) doi:10.1007/BF01807638.}

\bibitem{GQT1}{J.~Oliva, S.~Ray, ``A new cubic theory of gravity in five dimensions: black hole, Birkhoff's theorem and C-function,'' Class.Quant.Grav. 27  225002 (2010) doi:10.1088/0264-9381/ 27/22/225002 [arXiv:1003.4773 [gr-qc]].}

\bibitem{GQT2}{R.C.~Myers, B.~Robinson, ``Black Holes in Quasi-topological Gravity,'' JHEP 08 067 (2010) doi:10.1007/JHEP08(2010)067 [arXiv:1003.5357 [gr-qc]].}

\bibitem{TB}{C.~Teitelboim and J.~Zanelli, ``Gravitation theory generated by dimensional continuation of the Euler characteristic as a constrained Hamiltonian system, constraint’s theory and relativistic dynamics,'' Proceedings, Workshop, Florence, Italy, May 28- 30 (1986).}	

\bibitem{FOL}{N.~Deruelle, P.~Guilleminot, F.L.~Julie, N.~Merino, and R.~Olea, ``First-order Lagrangian and Hamiltonian of Lovelock gravity,'' Class.Quant.Grav. 38 10 (2021) doi:10.1088/1361- 6382/abf415 [arXiv:2011.01296 [gr-qc]].}
	
\bibitem{Pad2}{A.~Mukhopadhyay, T.~Padmanabhan, ``Holography of gravitational action functionals,'' Phys. Rev. D 74, 124023 (2006) doi:10.1103/PhysRevD.74.124023 [arXiv:hep-th/0608120 [hep-th]].}	
	
\bibitem{HL1}{H.~Khodabakhshi, H.~L\"u, ``Holographic relations and alternative boundary conditions for Lovelock gravity,'' Phys. Rev. D 95, [arXiv:2203.08839 [hep-th]].}


\bibitem{Lin}{D.~Glavan, C.~Lin,  ``Einstein-Gauss-Bonnet gravity in four-dimensional spacetime,'' Phys. Rev.Lett. 124  8, 081301 (2020) doi:10.1103/PhysRevLett.124.081301 [arXiv:1905.03601 [gr-qc]].}

\bibitem{LP}{H. L\"u and Y. Pang, ``Horndeski fravity as $D \rightarrow 4$ limit of Gauss-Bonnet,'' Phys.Lett.B 809 135717 (2020) doi:10.1016/j.physletb.2020.135717 [arXiv:2003.11552 [gr-qc]].}

\bibitem{TK}{T. Kobayashi, ``Effective scalar-tensor description of regularized Lovelock gravity in four dimensions,'' JCAP 07 013(2020) doi:10.1088/1475-7516/2020/07/013 [arXiv:2003.12771 [gr-qc]].}

\bibitem{Hennigar:2020lsl}
R.A.~Hennigar, D.~Kubiz\v{n}\'ak, R.B.~Mann and C.~Pollack,
``On taking the $D \rightarrow 4$ limit of Gauss-Bonnet gravity: theory and solutions,''
JHEP \textbf{07}, 027 (2020) [arXiv:2004.09472 [gr-qc]]

\bibitem{RR}{P.G.S.~Fernandes, P.~Carrilho, T.~Clifton, D.J.~Mulryne, ``The 4D Einstein-Gauss-Bonnet theory of gravity: a review,''  Class.Quant.Grav. 39 6, 063001 (2022) doi:10.1088/1361-6382/ac500a [arXiv:2202.13908 [gr-qc]].}

\bibitem{RB1}{R.B.~Mann and S.F.~Ross, ``The $D \rightarrow 2$ limit of general relativity,'' Class. Quant. Grav. 10 (1993) doi:10.48550/arXiv.gr-qc/9208004 [arXiv:9208004 [gr-qc]].}

\bibitem{Grumiller:2007wb}
D.~Grumiller and R.~Jackiw,
``Liouville gravity from Einstein gravity,''
[arXiv:0712.3775 [gr-qc]].


.\end{thebibliography}
\end{document}